\documentclass[12pt]{article}
\usepackage{epsfig}

\parskip=\medskipamount 
plus.3em minus.5em \abovedisplayshortskip=.5em plus.2em minus.4em
\renewcommand{\thesection}{\arabic{section}}

\catcode`@=11

\@addtoreset{equation}{section} \@addtoreset{equation}{subsection}
\def\theequation{\ifnum\value{section}=0 \arabic{equation}\ignorespaces
\else \ifnum\value{section}=-1 A.\arabic{equation}\ignorespaces
\else \ifnum\value{subsection}=0
\thesection.\arabic{equation}\ignorespaces \else
\thesection.\arabic{subsection}.\arabic{equation}\ignorespaces
                             \fi
                        \fi
                   \fi}

{\catcode`\'=\active \def'{{}^\bgroup\prim@s}}

\catcode`@=12

\newcommand{\bq}{\begin{equation}}
\newcommand{\be}{\begin{equation}}
\newcommand{\fq}{\end{equation}}
\newcommand{\ee}{\end{equation}}
\newcommand{\bqr}{\begin{eqnarray}}
\newcommand{\beqs}{\begin{eqnarray}}
\newcommand{\fqr}{\end{eqnarray}}
\newcommand{\eeqs}{\end{eqnarray}}

\newcommand{\rf}[1]{(\ref{#1})}

\def\bop#1{\setbox0=\hbox{$#1M$}\mkern1.5mu
    \vbox{\hrule height0pt depth.04\ht0
    \hbox{\vrule width.04\ht0 height.9\ht0 \kern.9\ht0
    \vrule width.04\ht0}\hrule height.04\ht0}\mkern1.5mu}

\begin{document}
\thispagestyle{empty}

\begin{flushright}
\begin{tabular}{l}
hep-th/0504219 \\
\end{tabular}
\end{flushright}

\vskip .6in
\begin{center}

{\bf  Tree Amplitudes in Gauge and Gravity Theories}

\vskip .6in

{\bf Gordon Chalmers}
\\[5mm]

{e-mail: gordon@quartz.shango.com}

\vskip .5in minus .2in

{\bf Abstract} 
\end{center} 

Gauge theory amplitudes in a non-helicity format are generated at all 
$n$-point and at tree level.  These amplitudes inherit structure from 
$\phi^3$ classical scattering, and the string inspired formalism is used 
to find the tensor algebra.  All of the classical gravity amplitudes 
are also given.  The classical effective action can also be constructed.  
Generalizations to amplitudes with non spin-$1$ or $2$ is possible.

\vfill\break 

\noindent{\it Introduction}

The tree amplitudes of gauge theories have recently been under much 
scrutiny, in view of the simplified derivation using the
weak-weak duality of the gauge theory with a twistor formulation.  The 
string inspired formulation of perturbative amplitudes, in addition to 
the techniques based on factorization and unitarity, have prompted 
further interest in their computation.  Helicity tree amplitudes 
up to $10$ point are presented in the literature, due to these methods.   

The gauge and gravity tree amplitudes are presented here at $n$-point.  
String inspired methods and scalar scattering are required to find 
their form.  The form can be used to find the full classical 
effective action in gauge theory and in gravity, in practical applications 
such as jet physics, and in further studies of duality.  An automation of 
the $n$-point tree amplitudes in gauge theory is useful for Monte Carlo 
simulations.

The massless $\phi^3$ diagrams and their specification are required 
in order to find the gauge theory and gravity theory tree-level 
amplitudes.  The string inspired representation of the latter 
utilizing the Koba-Nielsen amplitude, and its field theory limit, 
require the momentum routing of the propagators in a diagram and its 
correlated tensor algebra.  The form of the $\phi^3$ classical 
scattering presented in \cite{Chalmers1} has the required momentum 
parameterization.  The tensor algebra is computed with the scalar form.  

The gauge and gravity amplitudes are obtained in a no helicity basis, 
but in closed form via the known string-inspired tree-level rules.  
Spinor helicity is used in conjunction with the gauge invariance of 
the amplitudes to shorten the expressions.  The choice of reference 
momenta in the gauge invariant sets of diagrams that generates the 
simplest amplitude expression is not known in the literature; this 
algebraic question discussed in \cite{Chalmers2}, and its answer, is 
relevant for both formal and applied uses.  Different forms of the 
amplitudes lend to different interpretations and make manifest 
different properties, such as a twistor representation, a self-dual 
field form including a WZW model, or a potential iterative number basis 
form as in \cite{Chalmers2}.  The compact nature is important for 
practical applications. 

These classical amplitudes, both in scalar and gauge field theories, are 
required to bootstrap to higher orders.  The derivative expansion 
has been developed in \cite{Chalmers3}-\cite{Chalmers11} and the 
classical amplitudes are the initial conditions.

\vskip .2in 
\noindent {\it Gauge Amplitudes } 

The general scalar $\phi^3$ momentum routing of the propagators is 
presented.  Their form is required with the string scattering expression 
and the string inspired formulation to find the amplitudes.  

A general scalar field theory diagram at tree-level is parameterized 
by the set of propagators at the momenta labeling them.  In a color 
ordered form, consider the ordering of the legs as in $(1,2,\ldots,n)$.  
The graphs are labeled by 

\bqr 
D_\sigma = g^{n-2} \prod {1\over t_{\sigma(i,p)} - m^2} \ , 
\label{phi3diagrams} 
\fqr 
and the Lorentz invariants $t_{\sigma(i,p)}$ are defined by $t_i^{[p]}$, 
 
\bqr  
t_i^{[p]} = (k_i+\ldots + k_{i+p-1})^2 \ .  
\label{momentainv}
\fqr 
Factors of $i$ in the propagator and vertices are placed into the prefactor of 
the amplitude.
The sets of permutations $\sigma$ are what are required in order to specify 
the individual diagrams.  The full sets of $\sigma(i,p)$ form all of the 
diagrams, at any $n$-point order.  

The expansions in mass of a massive $\phi^3$ diagram follows from expanding 
the propagators, 

\bqr
{\cal A}^n_{\sigma,\tilde\sigma} = \sum_{\sigma,\tilde\sigma} 
 C_{\sigma\tilde\sigma} {g^{n-2}\over m^{n-2}} 
 \prod {t_{\sigma(i,p)}^{\tilde\sigma(i,p)}\over m^{2\tilde\sigma(i,p)}} \ .
\label{phi3nptclassicalmassive}
\fqr 
with the coefficient $C_\sigma$ determined from the momentum routing 
of the tree graphs; the $C$ coefficients take on non-empty values when 
there is a diagram.  An additional set of permutations $\tilde\sigma$ is 
required in order to specify the expansion of the propagators as in 
$(m^2-p^2)^{-1}=m^{-2}\sum (p^2/m^2)^l$.  

The massless diagrams are, 

\bqr
{\cal A}^n_\sigma = \sum_\sigma 
 C_\sigma g^{n-2} \prod t_{\sigma(i,p)}^{-1} \ , 
\label{phi3classicalmassless}
\fqr 
with the $C_\sigma$ spanning the set of all $(i,p)$ values at a given 
$n$-point.  The form and permutation set of $C_\sigma$ and 
$C_{\sigma\tilde\sigma}$ is given in \cite{Chalmers1}.

The vertices of the ordered $\phi^3$ diagram are labeled so that the outer numbers 
from a two-particle tree are carried into the tree diagram in a manner so that $j>i$ 
is always chosen from these two numbers.  The numbers are carried in from the $n$ 
most external lines. 

The labeling of the vertices is such that in a current or on-shell diagram the 
unordered set of numbers 
are sufficient to reconstruct the current; the set of numbers on the vertices are 
collected in a set $\phi_m(j)$.  For an $m$-point current there are 
$m-1$ vertices and hence $m-1$ numbers contained in $\phi_m(j)$.  These $m-1$ 
numbers are such that the greatest number may occur $m-1$ times, and must occur 
at least once, the next largest number may occur at most $m-2$ times (and may or 
may not appear in the set, as well as the subsequent ones), and so on.  The smallest 
number can not occur in the set contained in $\phi_m(j)$.  Amplitudes are treated 
in the same manner as currents.  Examples and a more thorough analysis is presented 
in \cite{Chalmers1}. 

Two example permutation sets pertaining to $4$- and $5$-point currents are:  

\bqr 
\pmatrix{ 444 \cr 
          443 \cr 
          442 \cr 
          433 \cr 
          432  }
 \label{threeparticle}
\fqr  

\bqr 
\pmatrix{ 5555 \cr 
          5554 \cr 
          5553 \cr 
          5552 \cr 
          5544 \cr 
          5543 \cr 
          5542 \cr 
          5( 3)}
\fqr 
with the $5(3)$ representing the $(3)$-permutation set attached to the $5$ in the 
total count.  There are $5$ and $15$ in the counts. The set of numbers in $\phi(j)$ 
is ordered from largest to least.

The numbers $\kappa(i)$ and $\phi(j)$ are used to find the propagators in the 
labeled diagram.  The procedure to determine the set of $t_i^{[p]}$, or the 
$\sigma(i,p)$, is as follows.  First, label all momenta as $l_i=k_i$.  Then, 
the invariants are found with the procedure,    

\vskip .2in 
1) $i=\phi(m-1)$, $p=2$, then $l_{a_{m-1}}+l_{a_m}\rightarrow l_{m-1}$ 

2) $i=\phi(m-2)$, $p=2$, then $l_{a_{m-2}}+l_{a_{m-1}}\rightarrow l_{m-2}$ 

... 

$m-1$) $i=1$, $p=m$
\bqr \label{sigmarules} 
\fqr 
\vskip .2in 

\noindent The labeling of the kinematics, i.e. $t_i^{[p]}$, is direct from the 
definition of the vertices.  

The numbers $\phi_n(i)$ can be arranged into the numbers $(p_i,[p_i])$, in which 
$p_i$ is the repetition of the value of $[p_i]$.  Also, if the number $p_i$ equals 
zero, then $[p_i]$ is not present in $\phi_n$.  These numbers can be used to 
obtain the $t_i^{[q]}$ invariants without intermediate steps with the momenta.  
The branch rules are recognizable as, for a single $t_i^{[q]}$,  

\vskip .2in
0) $l_{\rm initial}=[p_m]-1$

1) 

$r=1$ to $r=p_m$  

${\rm if~} r + \sum_{j=l}^{m-1} p_j = [p_m]-l_{\rm initial}   \quad {\rm then}~i= [p_m] 
  \quad q= [p_m] - l_{\rm initial}+1$ 

beginning conditions has no sum in $p_j$

2) 

${\rm else~} \quad l_{\rm initial}\rightarrow l_{\rm initial}-1$ : 
decrement the line number

$l_{\rm initial}>[p_{l}]$ else $l\rightarrow l-1$ : decrement the $p$ sum 

3) ${\rm goto}~ 1)$ 
\bqr  
\label{branchrules}
\fqr  
The branch rule has to be iterated to obtain all of the poles. 
This rule checks the number of vertices and matches to compare if there is a 
tree on it in a clockwise manner.  If not, then the external line number $l_{initial}$ 
is changed to $l_{initial}$ and the tree is checked again.  The $i$ and $q$ are labels 
to $t_i^{[q]}$.  

The previous recipe pertains to currents and also on-shell amplitudes.  There are 
$m-1$ poles in an $m$-point current $J_\mu$ or $m-3$ in an $m$-point amplitude.  The 
comparison between amplitudes and currents is as follows: the three-point vertex is 
attached to the current (in $\phi^3$ theory), and then the counting is clear when the 
attached vertex has two external lines with numbers 
less than the smallest external line number of the current (permutations to other 
sets of $\phi_n$ does not change the formalism).  There are $n-3$ poles are accounted 
for in the amplitude with $\phi_n$ and the branch rules.

\vskip .2in 
\noindent{\it Gauge Amplitudes: Tensor Algebra}

The gauge theory amplitudes are computed with the string inspired 
formulation.  The amplitudes are projected onto a color basis, so that 
only the amplitude with leg ordering $1,\ldots,n$ is analyzed.  The 
color prefactor is ${\rm Tr}T_{a_1}\ldots T_{a_n}$; permutations are 
used to obtain different orderings.  The momentum routing of the 
propagators, $C_\sigma$ is 
required in order to specify the combinations of products $\epsilon_i 
\cdot k_j$ and $\epsilon_i\cdot \epsilon_j$ appearing in a $\phi^3$ 
diagram.   The string-inspired formalism based on pinching the Koba-Nielsen 
formula generates a symmetric set of graph rules \cite{StringInspired} which 
is well adapted to the general expressions of scalar tree amplitudes.

The multi-linear string scattering expression is, 

\bqr 
\prod_{i\neq j} \exp{ \Bigl( ~k_i\cdot \varepsilon_j {\dot G}_B(i,j) 
 -{1\over 2} \varepsilon_i \cdot\varepsilon_j {\dot{\dot G}}_B(i,j) ~\Bigr) }  \ , 
\label{KNformula} 
\fqr 
in which the exponential is expanded to contain products of the 
polarization variables (i.e. multi-linear).  The world-sheet propagator 
term is 

\bqr 
\prod_{i\neq j} \exp{{1\over 2} k_i\cdot k_j ~G_B} \ , 
\label{worldsheet} 
\fqr   
which is useful in the field limit when the integration by parts is 
carried out on the ${\dot{\dot G}}_B$.  

The expansion of the polarizations in \rf{KNformula} generate the 
terms, 

\bqr 
\prod_{j=1}^n \left( \sum_{i\neq j} k_j\cdot \varepsilon_i 
  {\dot G}_B(i,j) \right)   
\fqr  
and 

\bqr  
\sum_\rho 
\prod_m^p \left(-{1\over 2}\right) 
  \varepsilon_{\rho(m,1;i)} \cdot \varepsilon_{\rho(m,2;i)} 
 ~ {\dot{\dot G}}_B(\rho(m,1;i),\rho(m,2;i))
\fqr 
\bqr 
\times \prod_{i\neq \rho(m,1;i),\rho(m,2;1)} 
  \left( \sum_{j\neq i} k_j\cdot \varepsilon_i {\dot G}_B(i,j) \right)
\label{multiexpansion} \ , 
\fqr 
which contains all possible products of the polarization vectors.  

The integration by parts on all of the ${\dot{\dot G}}_B(i,j)$ terms 
using the factor in \rf{worldsheet} generates the string-inspired form, 

\bqr 
\sum_\rho
\prod_m^p \left(-{1\over 2}\right) 
  \varepsilon_{\rho(m,1;a)} \cdot \varepsilon_{\rho(m,2;a)} 
 ~ {\dot G}_B(\rho(m,1;a),\rho(m,2;a))   
\fqr 
\bqr 
\times  
\hskip -.5in 
\left( {1\over 2} \sum_{c\neq b,b=\rho(m,2;a) } k_b \cdot k_c {\dot G}_B(b,c) \right)  
\prod_{i\neq \rho(m,1;i),\rho(m,2;1)} \quad
  \left( \sum_{j\neq i} k_j\cdot \varepsilon_i {\dot G}_B(i,j) \right)
\label{expansion} \ . 
\fqr 
The products in \rf{expansion} are used in conjunction with the individual 
$\phi^3$ diagrams to determine the gauge theory amplitudes.  The 
string-inspired rules, found in the field theory limit of the KN tree 
formula, are applied to the formula in \rf{expansion}.  

The permutation set $\rho(m,1;i)$ and $\rho(m,2;i)$ are derived via 
picking $p$ numbers from the collection $1,\ldots,n$ for each of the 
sets.  These $p$ numbers are non-overlapping in both $\rho(m,1;)$ and 
$\rho(m,2;)$ as there is no duplication of the $\varepsilon$ vectors.  
All possible combinations are required; there are $2 p!$ permutations 
of the sets' elements, and $n!/p!(n-p)!$ choices of the $p$ elements 
from the total number of $n$ polarizations.  The overcounting is 
$2^p$ as each of the pairs of polarizations is unordered.  

\begin{figure}
\begin{center}
\epsfxsize=12cm
\epsfysize=12cmhe
\epsfbox{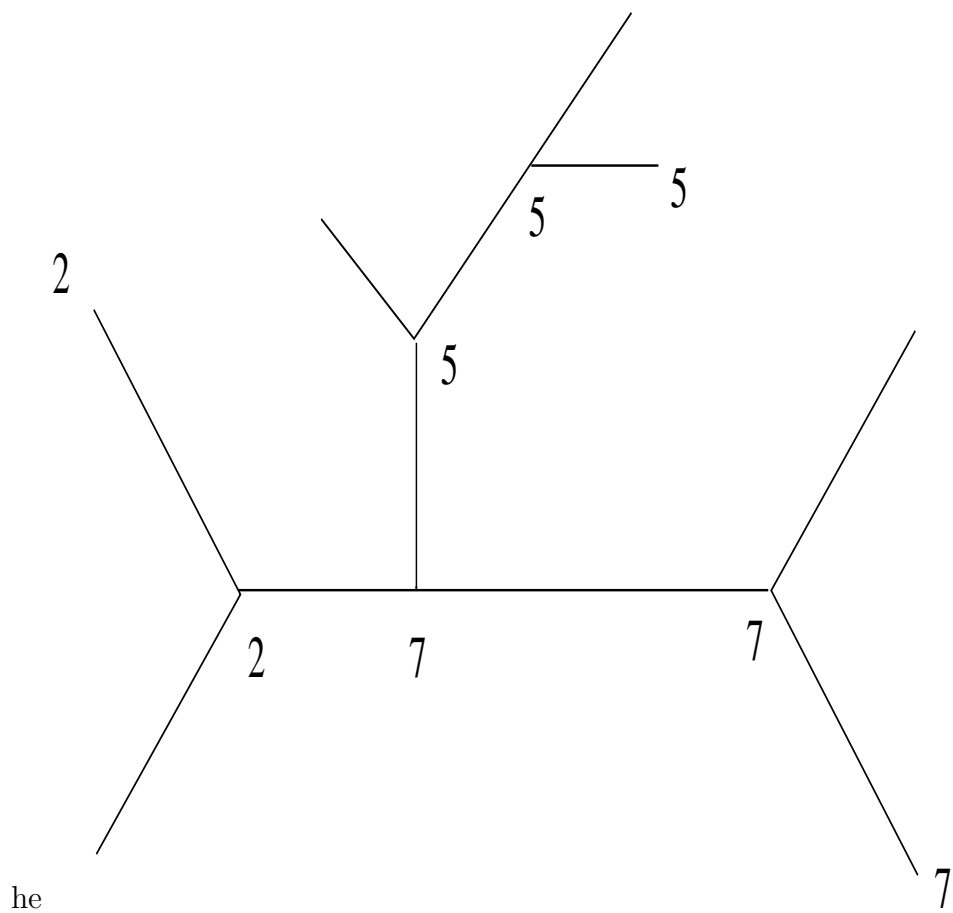}
\end{center}
\caption{The ordering and labeling of a sample $\phi^3$ diagram.}  
\end{figure}

The string inspired rules require that, from the expansion in \rf{expansion}, 
there are a certain number of ${\dot G}_B$ terms appearing in coordination 
with the labeling of the color ordered $\phi^3$ diagram.  The presence 
of the ${\dot G}_B$ terms is accompanied by the kinematic factors multiplying 
them, i.e. the tensor algebra. 
  
The vertices of the $\phi^3$ are numbered as in the previous section. 
The labeling of the numbers in the vertices is correlated with the set of 
numbers contained in $\sigma(i,p)$; physically, the specification 
of the multi-particle poles generates the labeling.  Each vertex number 
in a tree is labeled by taking the clockwise number in the two outer 
nodes $i$ and $j$ of the same tree.  The numbers are found by starting with the 
outermost external leg numbers, which range from $1$ to $n$ in a 
cyclic fashion for the color ordering $1,\ldots,n$.   The numbers 
are such that $i>j$, with $i$ and $j$ the two outer points on the two-particle 
tree, is always chosen from the choice of the two numbers.  

The vertices are associated with ${\dot G}_B(i,j)$ worldsheet bosonic 
Greens functions.  Each vertex requires one of these ${\dot G}_B$ with 
the two indices as: one of them labeled by the node, the other with 
one of the numbers so as to be in the outer tree.  
The choice of the latter to encompass the outer tree generates a 
'primary' choice of the ${\dot G}_B$.  Each vertex is associated with 
precisely one ${\dot G}_B$ factor; these factors are set equal to unity 
and the kinematics associated with the combination generates the 
tensor algebra. 

The set of numbers $\phi_n(i)$ discussed in the previous section gives 
a route to finding the poles $t_i^{[p]}$.  These numbers, as discussed 
in \cite{Chalmers1}, are representative of a discrete symmetry in the 
classical scattering.  The same numbers are used to find the tensor 
algebra on the scalar graphs of the gauge theory.

The primary set of vertex labels and their ${\dot G}_B$ factors are 
obtained from the momentum routing of the scalar diagram given 
in \rf{phi3classicalmassless}.  The collection of indices in 
$\sigma(i,p)$, which label $t_i^{[p]}$, generate the ${\dot G}_B$ 
factors with the indices $i,i+p-1$.   Denote this set of 
numbers as $\kappa(a,b)$ with $a$ and $b$ numbered by the propagator 
indices $i+p-1$ and $i$.  Beyond the primary set of indices, the 
remaining sets of indices have numbers with $i$ and $i+1,\ldots,i+p-1$, 
or $a$ to $a+1,\ldots, b$.  The numbers $i+p-1$ are identical to the 
$n-2$ vertex numbers in $\phi_n(i)$ (there are $n-2$ vertices and  
$n-3$ propagators and an overcount is on the last of the largest number 
in $\phi_n$).  The two sets of numbers are placed in the correlated 
$\kappa_1(i)$ and $\kappa_2(i)$

There are two more entries in $\kappa(a,)$ beyond the $n-2$ vertices; 
these entries are orthogonal numbers to the set contained in $\phi_n(i)$.  
The two entries in $\kappa(,i)$ beyond the $n-2$ primary and 
descendent numbers ($i$ to $i+1,\ldots,i+p-1$) and comprise an orthogonal 
set of ${\dot G}_B(i,j)$.  
   
The vertex labels are used to extract $n-2$ non-identical ${\dot G}_B(i,j)$ 
factors.  These factors are pulled from the kinematic expression in 
\rf{expansion}.  The remaining factors must have non-overlapping 
${\dot G}_B(i,j)$; all of the bosonic Greens functions are then 
set to $1$ or $-1$ for $i>j$ or $j<i$.  

The $\kappa(a;1)$ and $\kappa(b;2)$ set of primary numbers used on 
\rf{expansion} produces a contribution, 

\bqr 
(-{1\over 2})^{a_1} ({1\over 2})^{n-a_2} 
\prod_{i=1}^{a_1} \varepsilon(\kappa(i;1))\cdot \varepsilon(\kappa(i;1))  
\times \prod_{j=a_1+1}^{a_2} \varepsilon(\kappa(j;1)) \cdot k_{\kappa(j;2)} 
\times \prod_{p=a_2+1}^n k_{\kappa(p;1)} \cdot k_{\kappa(p;2)}  \ ,
\label{kappaterms}
\fqr 
together with the permutations of $1,\ldots,n$.  The permutations extract 
all possible combinations from the \rf{expansion}, after distributing the 
numbers into the three categories.   

The form of the amplitudes are expressed as, 

\bqr
{\cal A}^n_\sigma = \sum_\sigma 
 C_\sigma g^{n-2} T_\sigma \prod t_{\sigma(i,p)}^{-1} \ , 
\label{gaugeamps}
\fqr 
with $T_\sigma$ in \rf{kappaterms} derived from the tensor set of 
$\kappa$, e.g. found from $\phi_n$ or the momentum routing of the propagators 
with $\sigma(i,p)$.   The normalization is $i(-1)^n$.  The numbers 
$a_1$ and $a_2$ are summed so 
that $a_1$ ranges from $1$ to $n/2$, with the boundary condition 
$a_2\geq a_1+1$.  Tree amplitudes in gauge theory must possess at least 
one $\varepsilon_i\cdot\varepsilon_j$.

All $\phi^3$ diagrams are summed at $n$-point, which is represented by the 
sum in $\sigma$ in \rf{gaugeamps}.  The color structure is 
${\rm Tr} \left(T_{a_1}\dots T_{a_n}\right)$,  and the complete amplitude 
involves summing the permutations of $1,\ldots, n$.  

The first $n-2$ numbers in $\kappa_2$ are summed beyond those of the primary 
numbers in accord with the set $i$ to $i+p-1$ for a given vertex label 
$i+p-1$, which labels the vertex in $\phi_n$.   

\vskip .2in 
\noindent {\it Gravity Amplitudes}

Graviton scattering is straightforward given the gauge theory results.  
The holomorphic gauge theory string derivation is squared, i.e. the 
tensor algebra must include an identical non-holomorphic piece.  
The multi-linear string scattering expression is, 

\bqr 
\prod_{i\neq j} \vert \exp{ \Bigl( ~k_i\cdot \varepsilon_j {\dot G}_B(i,j) 
 -{1\over 2} \varepsilon_i \cdot\varepsilon_j {\dot{\dot G}}_B(i,j) ~\Bigr) } 
 \vert^2  \ , 
\label{KNformulasquare} 
\fqr 
and contains the holomorphic square of the function in \rf{KNformula}.

The world-sheet propagator term is squared 

\bqr 
\prod_{i\neq j} \exp{{1\over 2} k_i\cdot k_j (G_B+{\bar G}_B)} \ . 
\label{worldsheetsquare} 
\fqr   
The integration by parts on all of the ${\dot{\dot G}}_B(i,j)$ terms 
using the factor in \rf{worldsheet} includes the product of the 
non-holomorphic half of the string-inspired form, i.e. the barred 
piece of \rf{expansion}.

The gravitational amplitudes are, via the holomorphic splitting, 
\bqr
{\cal A}^n_\sigma = \sum_\sigma 
 C_\sigma g^{n-2} T_\sigma {\bar T}_\sigma 
 \prod t_{\sigma(i,p)}^{-1} \ , 
\label{gravitytrees}
\fqr 
with a holomorphically squared $T_\sigma$, the same as in gauge theory,  

\bqr 
T_\sigma = \sum_{\mu,\nu,\gamma} 
 {\tilde T}_{\mu\nu\gamma}  
  \prod_i k_{\mu(i;1)}\cdot \varepsilon_{\mu(i;2)}  
  \prod_j \varepsilon_{\nu(j;1)} \cdot \varepsilon_{\nu(j;1)}  
           \prod_s k_{\gamma(s;1)} \cdot k_{\gamma(s;2)} \ .
\label{numerator} 
\fqr 
This form is the complete gravitational S-matrix, after summing the 
orderings of the external legs.  

Both the gauge and gravitational scatterings may be gauge covariantized 
using the field strengths $F_{\mu\nu}$ and $R_{\mu\nu}$, to write the 
classical effective action.  The classical effective action is relevant 
to the DBI work, soliton effects including black hole dynamics, and 
the anti-de Sitter correspondence with gauge theory. 

\vskip .2in 
\noindent {\it Concluding remarks} 

Gauge and gravity amplitudes are found at tree level and with any number of 
legs.  The analagous scalar field theory amplitudes have appeared in 
\cite{Chalmers1}.  The amplitudes are generated without specifying the 
helicity content.  Two sets of numbers are required to delimit the contributions, 
$\phi_n$ of the vertices and $t_i^{[p]}$ of the poles, and they are equivalent.  

Gravity interactions, by varying $\int d^dx R\sqrt{g}$, possess an infinite 
number of vertices and makes the calculation of an all $n$-point formula tedious.  
The amplitudes are found by utilizing the KLT factorization of string tree scattering.  

These gauge and gravity amplitudes use the number parameterization of $\phi^3$ 
diagrams.  The latter have 'symmetries' accorded to them via the collection of 
numbers $\phi_n(j)$ used to construct the individual graphs (see \cite{Chalmers1}).  
These sets may also be used to classically quantize gauge and gravity, which 
should generalize to the quantum level.  

The string form of the scattering amplitudes together with the scalar field 
results should allow for a generalization of the all $n$ amplitudes to contain 
fermionic modes, $(p,q)$ tensor modes, mixed spins in the asymptotic states, 
and including the electroweak sector.

These amplitudes are required in order to bootstrap to the quantum level.  Also, 
a closed form of the classical amplitudes is useful for Monte Carlo simulations 
for particle beam simulations.

\vfill\break 
 
\end{document}